\documentclass[aip,jap,a4,reprint,twoside,floatfix,superscriptaddress]{revtex4-1}

\usepackage{epsfig}                                                            
\usepackage{graphicx}
\usepackage{dcolumn}
\usepackage{color}
\usepackage{bm}
\usepackage{subfigure}
\usepackage{float}

\begin{document}

\title {\bf Room-temperature multiferroicity in GaFeO$_3$ thin film grown on (100)Si substrate}

\author {Sudipta Goswami} \email {drsudiptagoswami@gmail.com} \address {School of Materials Science and Nanotechnology, Jadavpur University, Kolkata 700032, India}
\author {Shubhankar Mishra} \address {School of Materials Science and Nanotechnology, Jadavpur University, Kolkata 700032, India}
\author {Kausik Dana} \address {Refractories and Traditional Ceramics Division, CSIR-Central Glass and Ceramic Research Institute, Kolkata 700032, India}
\author {Ashok Kumar Mandal} \address {Materials Characterization and Instrumentation Division, CSIR-Central Glass and Ceramic Research Institute, Kolkata 700032, India}
\author {Nitai Dey} \address {Materials Characterization and Instrumentation Division, CSIR-Central Glass and Ceramic Research Institute, Kolkata 700032, India}
\author {Prabir Pal} \address {Materials Characterization and Instrumentation Division, CSIR-Central Glass and Ceramic Research Institute, Kolkata 700032, India}
\author {Biswarup Satpati} \address {Surface Physics and Materials Science Division, Saha Institute of Nuclear Physics, A CI of Homi Bhabha National Institute, 1/AF Bidhannagar, Kolkata 700064, India}
\author {Mrinmay Mukhopadhyay} \address {Surface Physics and Materials Science Division, Saha Institute of Nuclear Physics, A CI of Homi Bhabha National Institute, 1/AF Bidhannagar, Kolkata 700064, India}
\author {Chandan Kumar Ghosh} \address {School of Materials Science and Nanotechnology, Jadavpur University, Kolkata 700032, India}
\author {Dipten Bhattacharya} \address {Advanced Materials and Chemical Characterization Division, CSIR-Central Glass and Ceramic Research Institute, Kolkata 700032, India}

\date{\today}

\begin{abstract}
Room-temperature magnetoelectric multiferroicity has been observed in c-axis oriented GaFeO$_3$ thin films (space group $Pna2_1$), grown on economic and technologically important (100)Si substrates by pulsed laser deposition technique. Structural analysis and comprehensive mapping of Ga:Fe ratio across a length scale range of 10$^4$ reveal coexistence of epitaxial and chemical strain. It induces formation of finer magnetic domains and large magnetoelectric coupling - decrease in remanent polarization by $\sim$21\% under $\sim$50 kOe. Magnetic force microscopy reveals presence of both finer ($<$100 nm) and coarser ($\sim$2 $\mu$m) magnetic domains. Strong multiferroicity in epitaxial GaFeO$_3$ thin films, grown on (100)Si substrate, brighten the prospect of their integration with Si-based electronics and could pave the way for development of economic and more efficient electromechanical, electrooptic or magnetoelectric sensor devices.
\end{abstract}

\maketitle

\section{Introduction}

\begin{figure*}[ht!]
\begin{center}
   \subfigure[]{\includegraphics[scale=2.00]{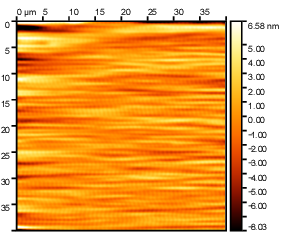}} 
   \subfigure[]{\includegraphics[scale=0.40]{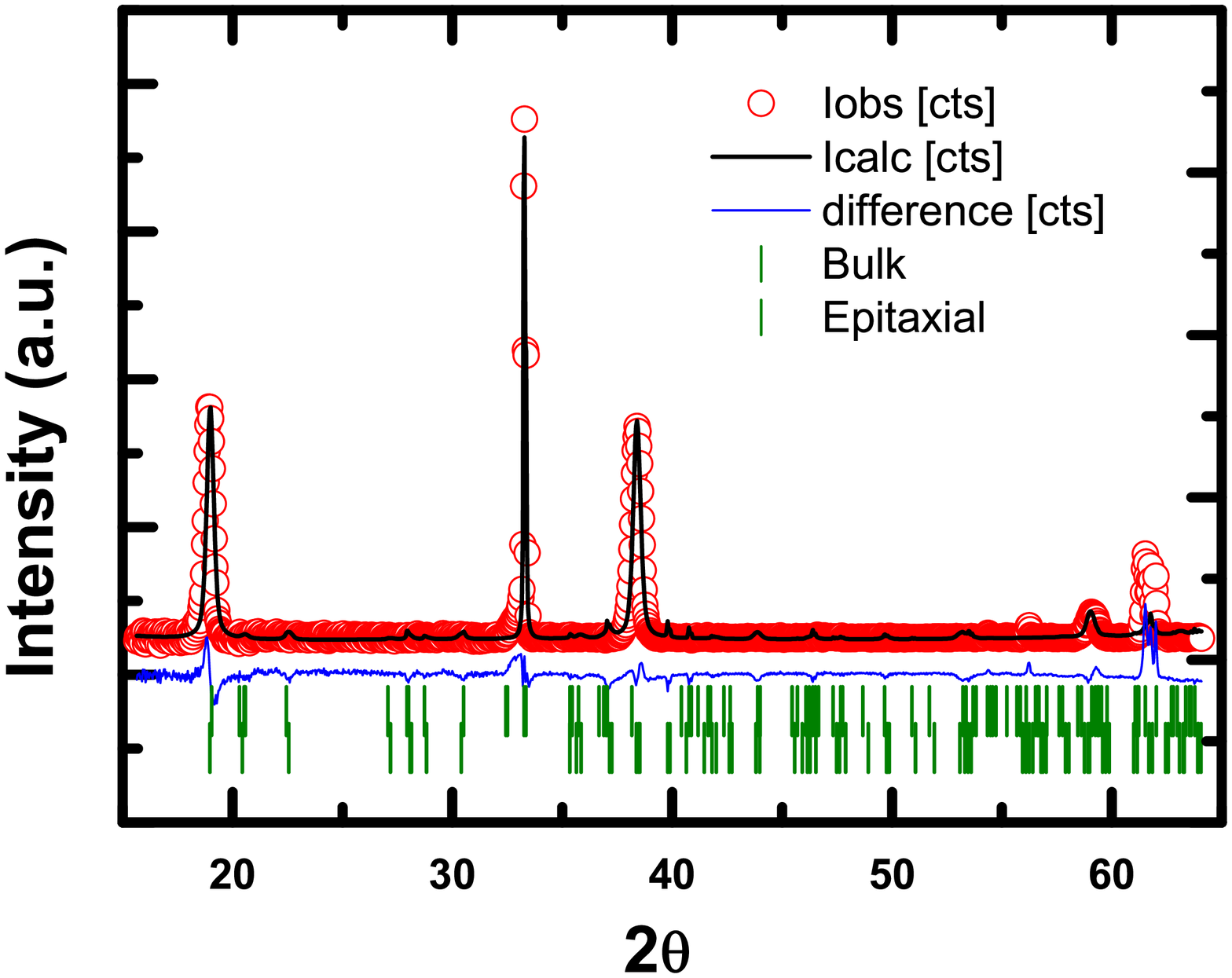}}
   \end{center}
\caption{(a) The atomic force microscopy (AFM) image of the surface of the GaFeO$_3$ film; (b) the grazing incidence x-ray diffraction (GIXRD) data for the film and their Rietveld refinement. }
\end{figure*}

Bulk GaFeO$_3$ (with Ga:Fe ratio = 1.0) exhibits robust ferroelectricity\cite{Mishra-1} till 1368 K due to off-centering of Fe1 and Fe2 ions along b-axis (within $Pc2_1n$ structure) and ferrimagnetism along c-axis below $T_N$ $\approx$ 220 K due to inequivalence\cite{Mohn} of Fe1 and Fe2 ions. Magnetoelectric multiferroicity, therefore, is observed below room temperature. Increase in Fe ion concentration $x$ across the composition range Ga$_{2-x}$Fe$_x$O$_3$ (0.8$\le$x$\le$1.4) gives rise to increase in $T_N$ from $\sim$100 to $\sim$350 K.\cite{Arima} The disorder in the Ga and Fe site occupancy\cite{Mohn} too influences the $T_N$. The linear magnetoelectric coupling originates here from additional polarization induced by magnetic ordering ($\Delta P$ = $P_{structural}$ $\pm$ $P_{magnetic}$ [Ref. 3]). In recent time, multiferroicity in GaFeO$_3$ has been explored by using different platforms - single crystals, epitaxial thin films, different nanostructures. Room temperature multiferroicity has been observed\cite{Niu} in polar corundum structure synthesized under high pressure and temperature. Room temperature multiferroicity has also been claimed in nanofibers\cite{Zhao} ($\textit{albeit}$ in an Fe-rich composition) and in epitaxial thin films\cite{Mukherjee} grown on cubic (001)YSZ substrate. Role of epitaxial strain and disorder in the site occupancy was examined by growing films primarily on cubic or hexagonal YSZ, Al$_2$O$_3$, and SrTiO$_3$ single crystal substrates.\cite{Mukherjee,Thai,Kundaliya,Trassin,Roy,Zhang,Sharma,Itoh,Song,Dugu} Formation of polar and magnetic nanodomains could be responsible for the observations in thin films. Of course, room-temperature multiferroicity has not been observed in all the cases. Very limited work, however, has so far been done\cite{Mishra-2} on the use of (100)Si substrate. In those cases, the films were prepared, primarily, by sol-gel technique. 

Epitaxial multiferroic thin films grown on Si substrate are extremely useful because of the possibility of their integration with the well-developed Si-based electronic device technology. Not only Si substrates are cheaper, they offer time-tested platforms for the entire range of micro- or nanoelectronic and spintronic devices. Therefore, integration of thin films of functional oxides with silicon has been an active area of research for many years now.\cite{Ghosez} Successful growth of ferroelectric, magnetic, or multiferroic oxide thin films on silicon substrate offers the potential of fabrication of the complete device package comprising of other functional elements such as FET etc.

Given this backdrop, we report here growing GaFeO$_3$ epitaxial thin film by pulsed laser deposition technique on economic and technologically important (100)Si substrate and show that the coexistence of strain (due to lattice mismatch between the substrate and the film) and $\textit{inhomogeneity}$ in the distribution of Ga:Fe ratio across the film over a length scale range of 10$^4$ induces room temperature multiferroicity. The film (thickness $\approx$ 70 nm) turns out to be preferentially oriented along c-axis with $\sim$94\% epitaxy. While, in general, persistence of ferro orders could be observed beyond a critical thickness,\cite{Fong} recent investigations\cite{Ju,Yuan,Homkar} reveal presence of long-range orders even down to the ultrathin limit (i.e. in films of thickness of the order of 1-10 nm or one to few unit cells). The extent of epitaxial strain and defect/dislocation concentration, of course, varies\cite{Jiang} with film thickness depending on the misfit strain between the substrate and the film. This, in turn, gives rise to thickness dependent multiferroicity. We, of course, focused here on the multiferroicity in GaFeO$_3$ thin film of $\sim$70 nm thickness on technologically relevant Si(100) susbtrate. Significant suppression of dielectric constant and ferroelectric polarization under magnetic field (0-50 kOe) provides clear evidence of room-temperature magnetoelectric multiferroicity.

\section{Experimental}
The thin films of orthorhombic GaFeO$_3$ were grown on (100)Si substrates (containing p-type carriers) by pulsed laser deposition technique using KrF excimer laser ($\lambda$ = 248 nm) of energy $\sim$1.20 J/cm$^2$ in a chamber maintained at $\sim$5.5$\times$10$^{-3}$ Torr pressure. The substrate temperature was $\sim$650$^o$C. We focus on a film grown by irradiating the target with 10000 laser pulses. The film was subsequently annealed for 30 minutes under oxygen before cooling down to room temperature. The atomic force microscopy (AFM) was used for determining the surface features and the thickness while the $\theta$-2$\theta$ as well as grazing incidence x-ray diffraction (GIXRD) scans (at a grazing angle 0.1$^o$) were carried out to extract the structural details. The field emission scanning electron microscopy (FESEM) along with energy dispersive x-ray spectra (EDX) were recorded to map out the composition at different length scale. In addition, different (primarily, in-plane) regions of the film were exposed for mapping the local lattice strain within a length scale of a few nanometers by transmission electron and high resolution transmission electron microscopy (TEM and HRTEM). The X-ray photoelectron spectroscopy (XPS) measurements were performed by PHI 5000 VERSAPROBE II, Physical Electronics System, equipped with monochromatic Al $k_{\alpha}$ (1486.7 eV) focussed x-ray source and a multi-channeltron hemispherical electron energy analyzer. All the spectra were collected at an emission angle of 45$^o$ with the base vacuum of 5.0 $\times$ 10$^{-10}$ mbar. The binding energies were referenced by measuring C 1s and keeping it at 284.6 eV. The total energy resolution was estimated to be $\sim$400 meV for monochromatic Al $k_{\alpha}$ line with pass energy 11.750 eV. A charge neutralizer was used to compensate the surface charging of the samples. A background has been subtracted from the measured raw data. The dielectric and ferroelectric properties were measured by using two-probe top-top electrode configuration with room temperature curable Ag electrodes by, respectively, the ferroelectric loop tester (Radiant Technologies Inc., Precision LC-II) and the impedance analyzer (IM3570, Hioki). The magnetization was measured in a SQUID magnetometer (Quantum Design) across 5-390 K under 500 Oe field applied parallel to the film surface. The magnetic force microscopy (MFM) was also employed using the LT-AFM/MFM System of Nanomagnetics Instruments Ltd., Ankara, Turkey, for imaging the room temperature magnetic domain structure of the film under zero magnetic field. The atomic force microscopy (AFM) imaging of the film surface features was carried out in the tapping mode by using a Michelson type interferometer detector. The cantilever (PointProbePlus Magnetic Force Microscopy - Reflex coating) was oscillated at the resonant frequency ($\sim$70 kHz) by a digital phase-lock-loop (PLL) control system. The cantilever tip-surface interaction was monitored by the rms value of the oscillation voltage which was kept constant at $\sim$0.5 V during the scan. This ensures stable interaction between the tip and the surface. Influence of mechanical and electrical noise was eliminated.

\begin{figure*}[t]
\centering
{\includegraphics[scale=0.50]{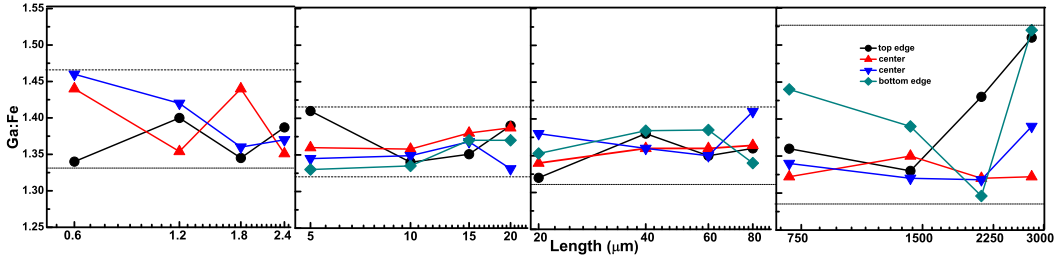}}
\caption{The mapping of the Ga:Fe ratio across different length scales - from macroscopic (over an area of mm$^2$) to nanoscopic (over an area of nm$^2$) }
\end{figure*}

\begin{figure*}[ht!]
\begin{center}
   \subfigure[]{\includegraphics[scale=0.20]{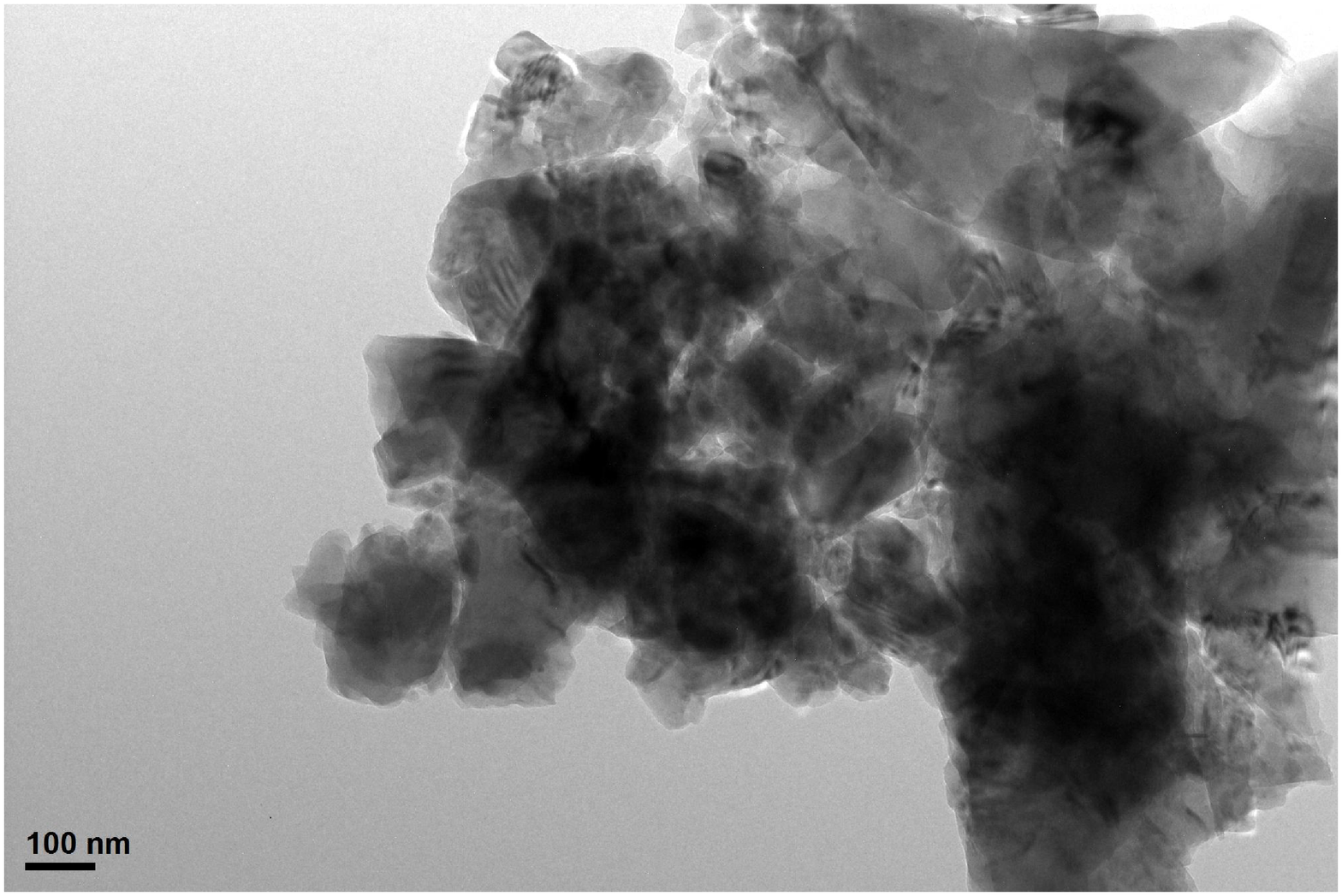}} 
   \subfigure[]{\includegraphics[scale=0.20]{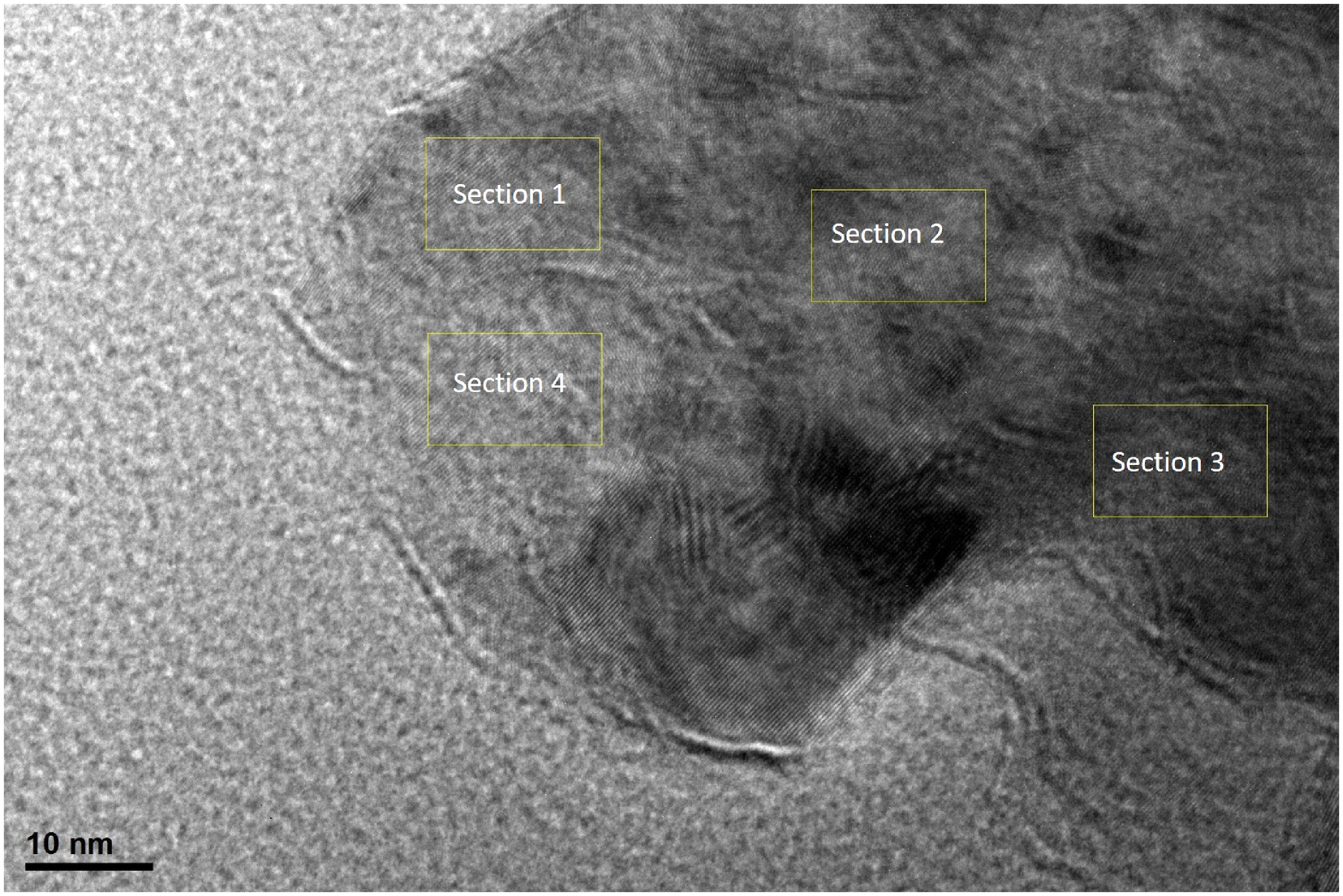}}
	\subfigure[]{\includegraphics[scale=0.20]{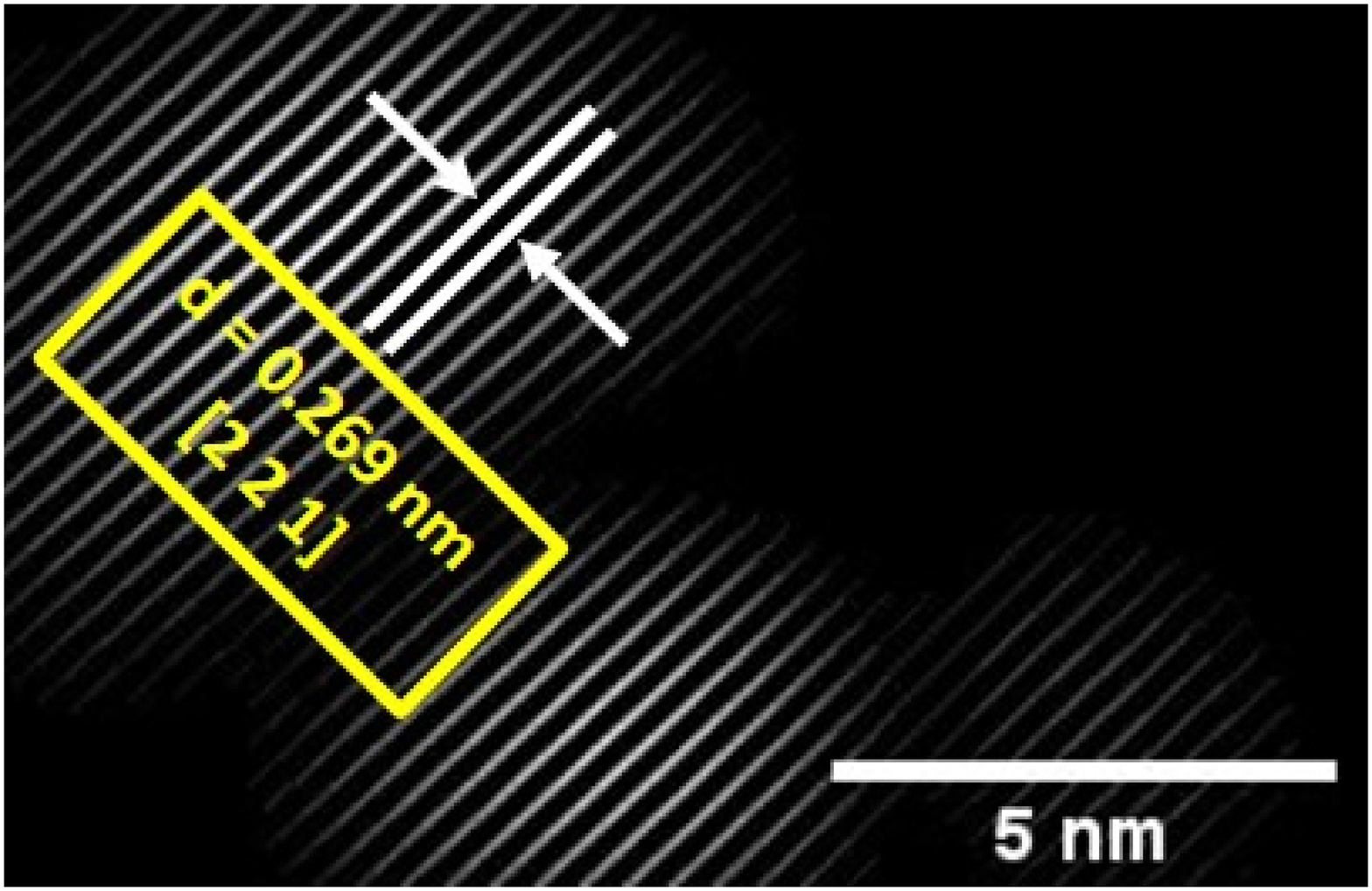}}
   \subfigure[]{\includegraphics[scale=0.20]{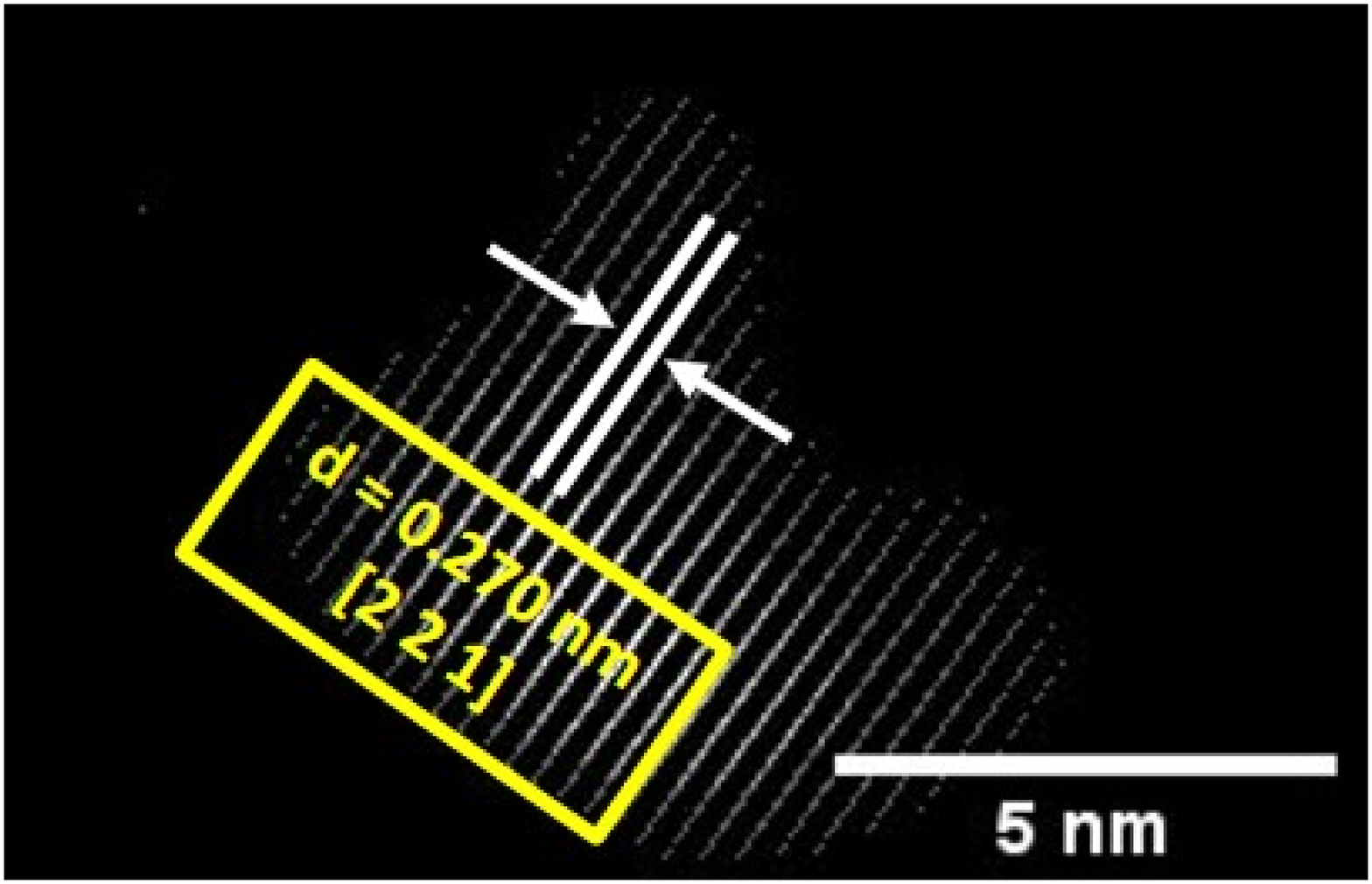}} 
   \subfigure[]{\includegraphics[scale=0.20]{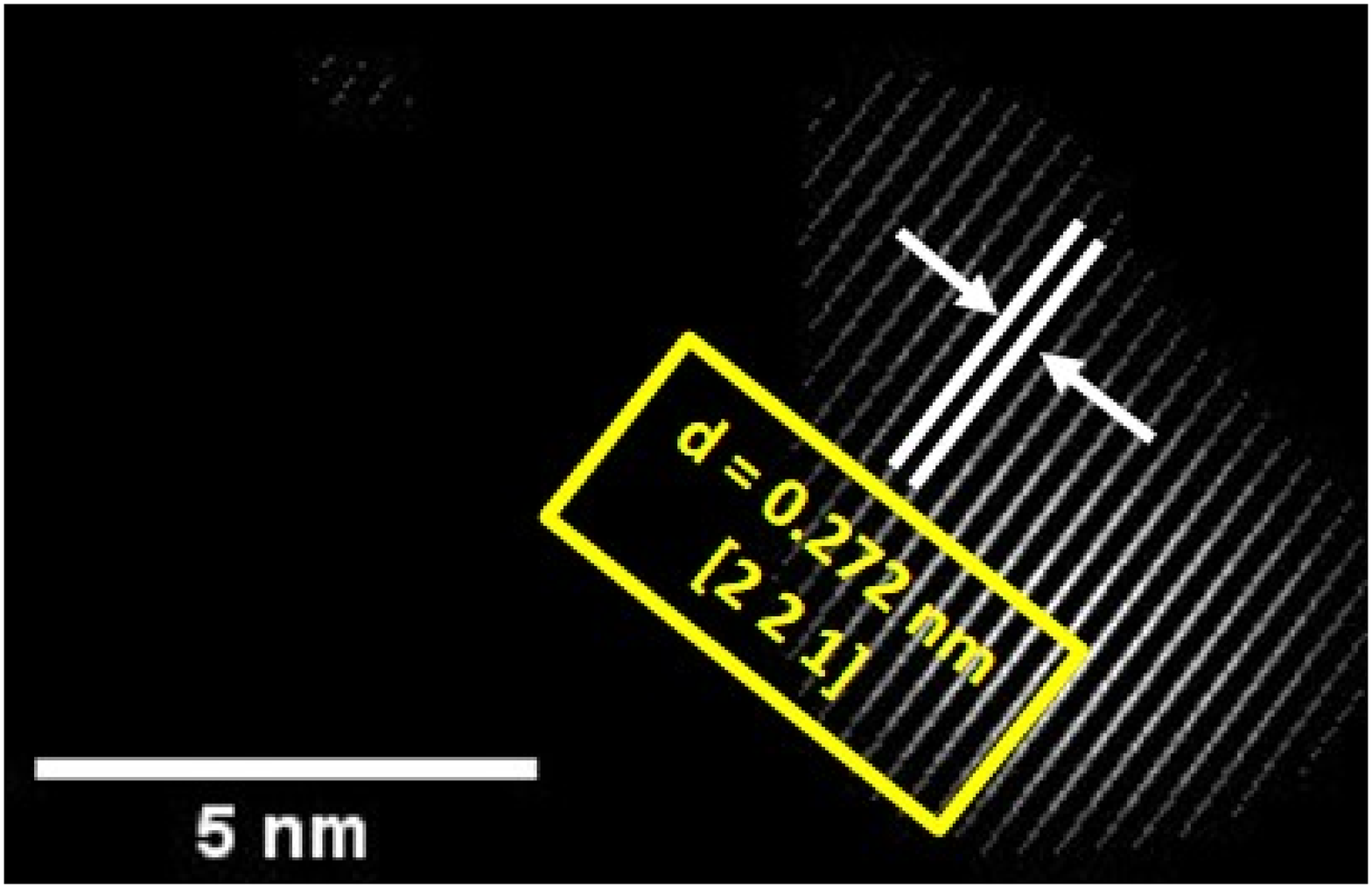}}
	\subfigure[]{\includegraphics[scale=0.20]{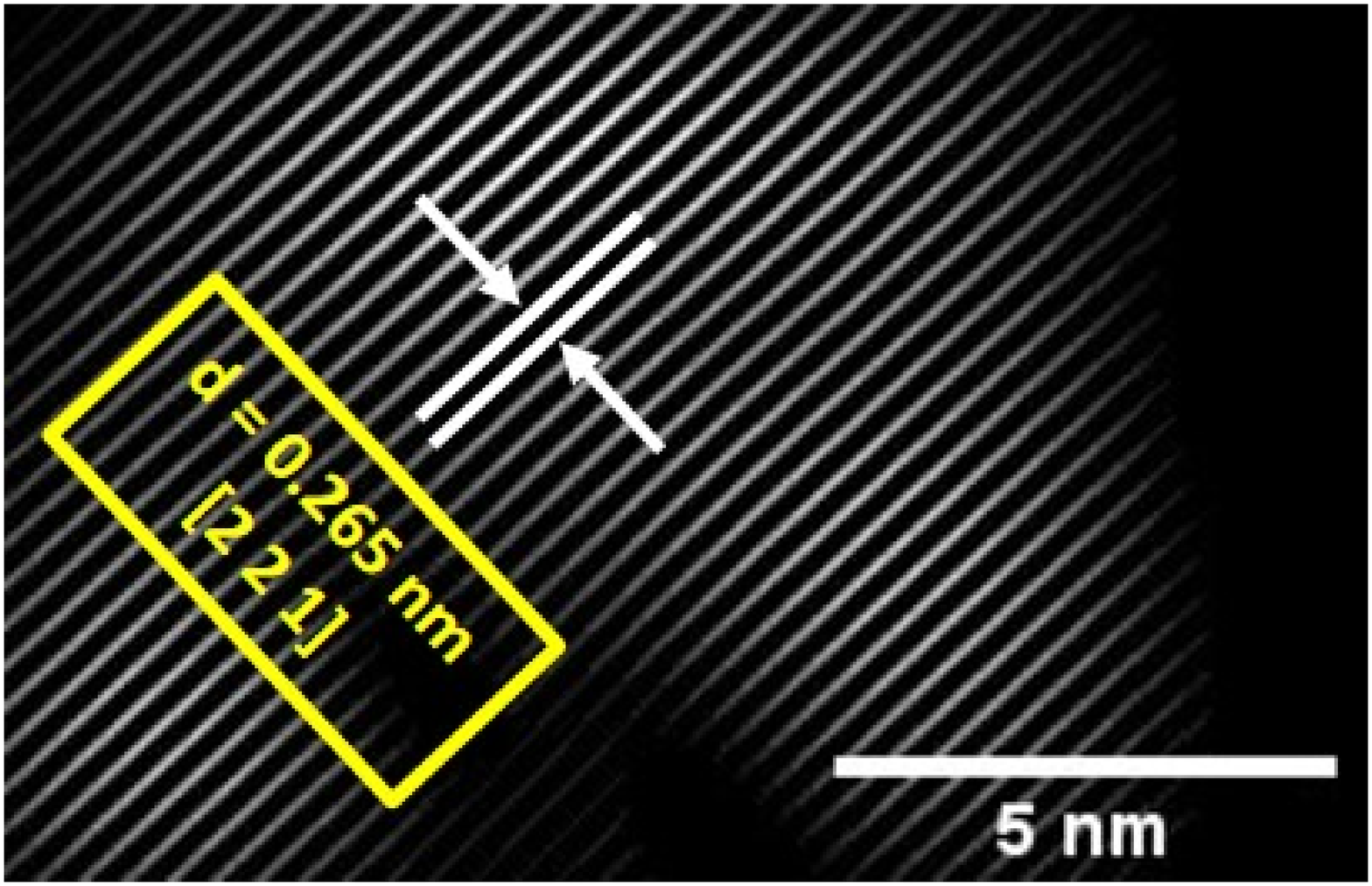}}
   \end{center}
\caption{(a) Representative transmission electron microscopy (TEM) image of the GaFeO$_3$ film and (b) the corresponding high resolution TEM (HRTEM) image; (c), (d), (e), (f) show, respectively, the analyzed lattice fringe structures corresponding to the boxes marked as sections 1-4 in (b); variation in the lattice spacing `$d$' for the plane (221) could be observed.    }
\end{figure*}

\section{Results and Discussion}
In Fig. 1(a) we show the AFM image of the film surface. The thickness is found to be $\sim$70 nm (surface roughness $R_a$ $\approx$ 2.74 nm). The Rietveld refinement of the bulk sample XRD data (supplementary material) yields the lattice parameters to be a = 5.089(5) \AA, b = 8.693(6) \AA, and c = 9.366(4) \AA, for $Pna2_1$ structure. The $\theta$-2$\theta$ XRD scan (supplementary material) shows preferential growth of the film along c-axis. The (001) plane, i.e., the ab-plane, of GaFeO$_3$ grows onto the (100) plane of the Si substrate (lattice constant of the Si substrate is 5.410 \AA). The GIXRD data (Fig. 1b), on the other hand, point out textured nature of the film (i.e., neither purely epitaxial nor polycrystalline) as presence of other peaks could also be observed. Refinement of the GIXRD data shows conformation with the $Pna2_1$ structure with lattice parameters a = 5.050(1) \AA, b = 8.680(2) \AA, and c = 9.404(3) \AA. Quantitative analysis of the data, where integrated intensity of the (002), (004), (006), and (008) peaks and that of the entire background were determined, yields the extent of preferential orientation along the c-axis to be nearly $\sim$94\%. Using the lattice parameters of the bulk target and the thin film, the lattice strain along c-axis and within ab-plane ($s_{\perp}$, $s_{\parallel}$) is estimated to be $s_{\perp}$ $\approx$ +0.13\%, $s_{\parallel}$ $\approx$ -0.69\%. The strain is, therefore, slightly tensile along c-axis and compressive within the ab-plane.

\begin{figure}[ht!]
\begin{center}
   \subfigure[]{\includegraphics[scale=0.20]{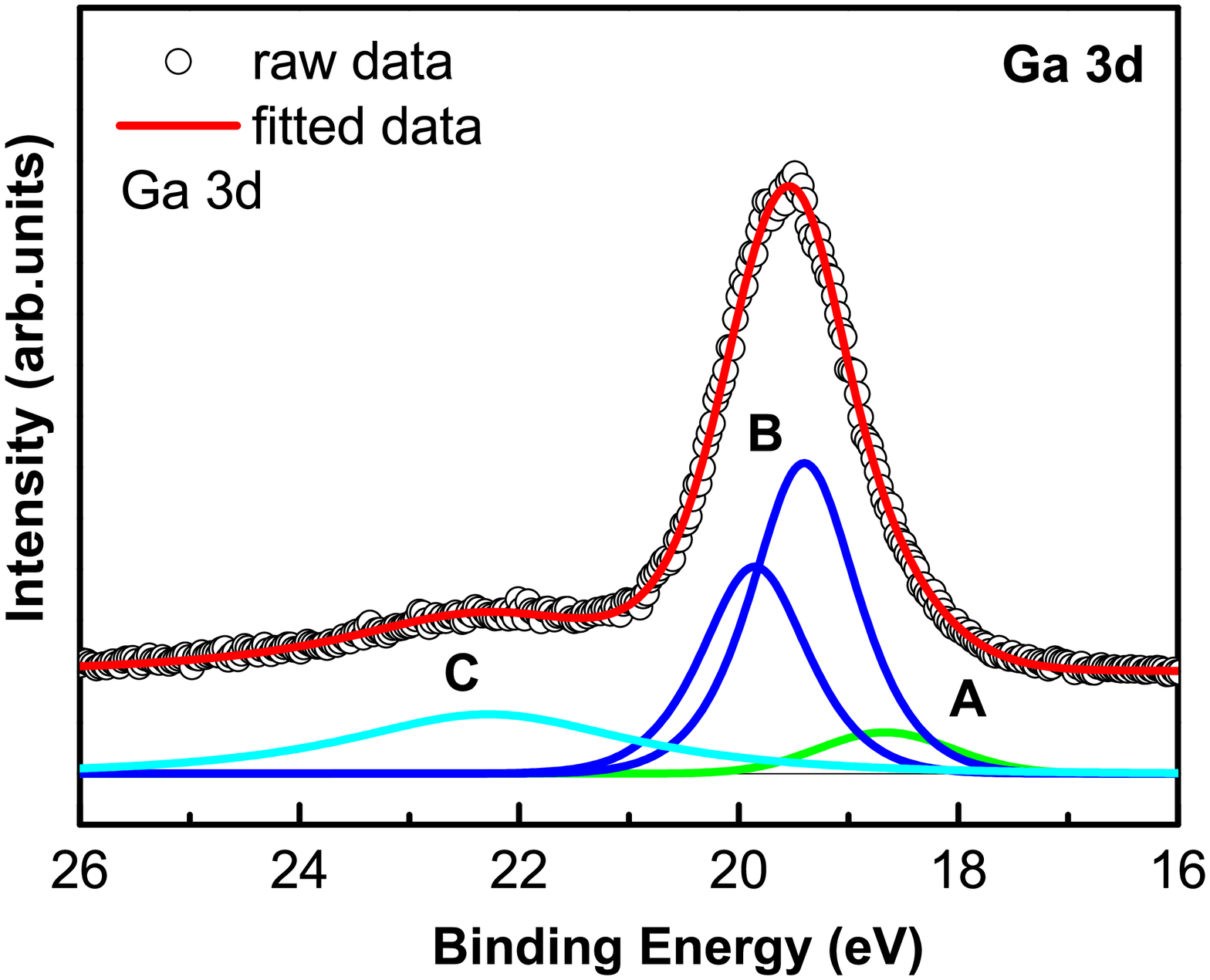}} 
   \subfigure[]{\includegraphics[scale=0.20]{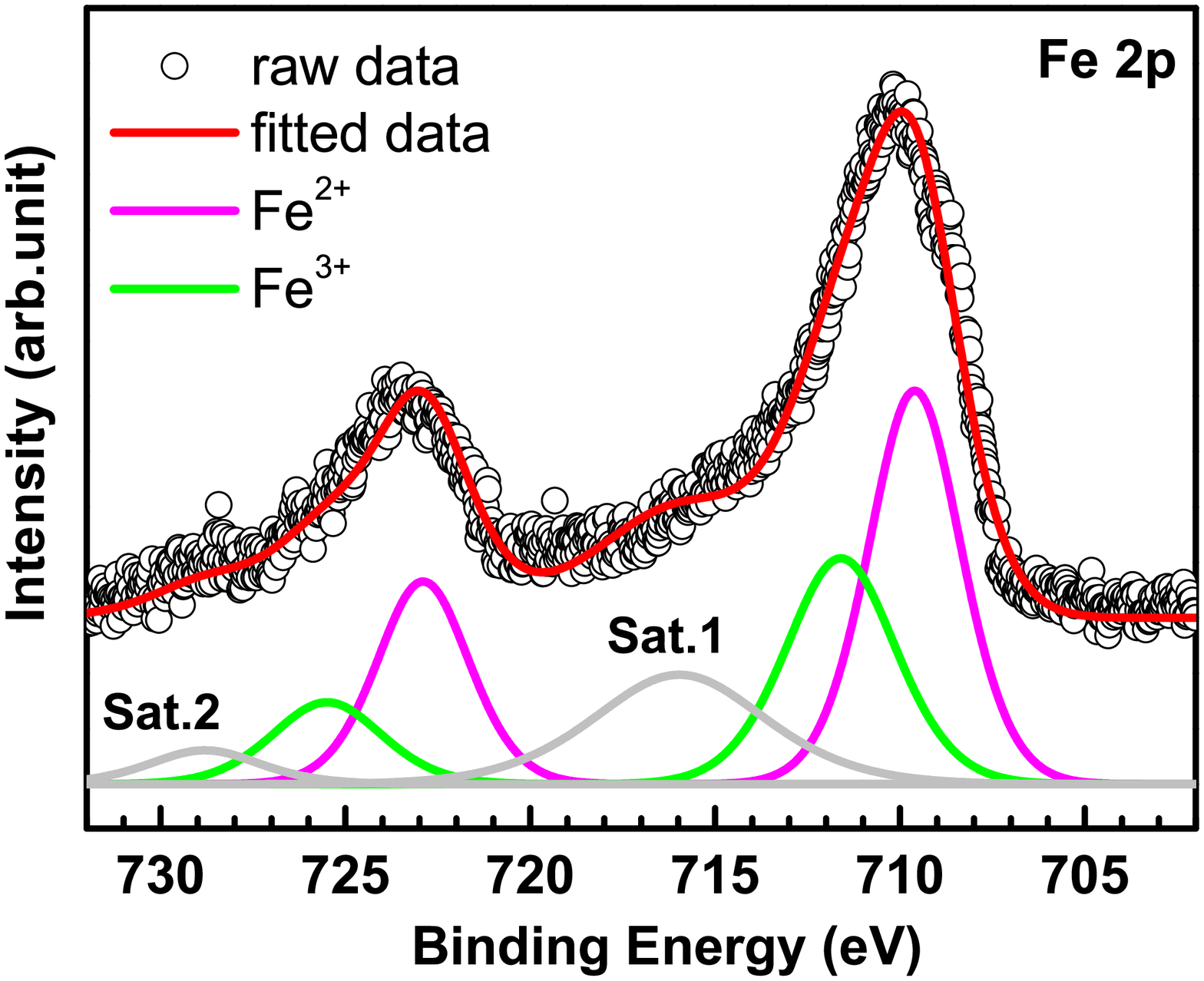}}
	\subfigure[]{\includegraphics[scale=0.20]{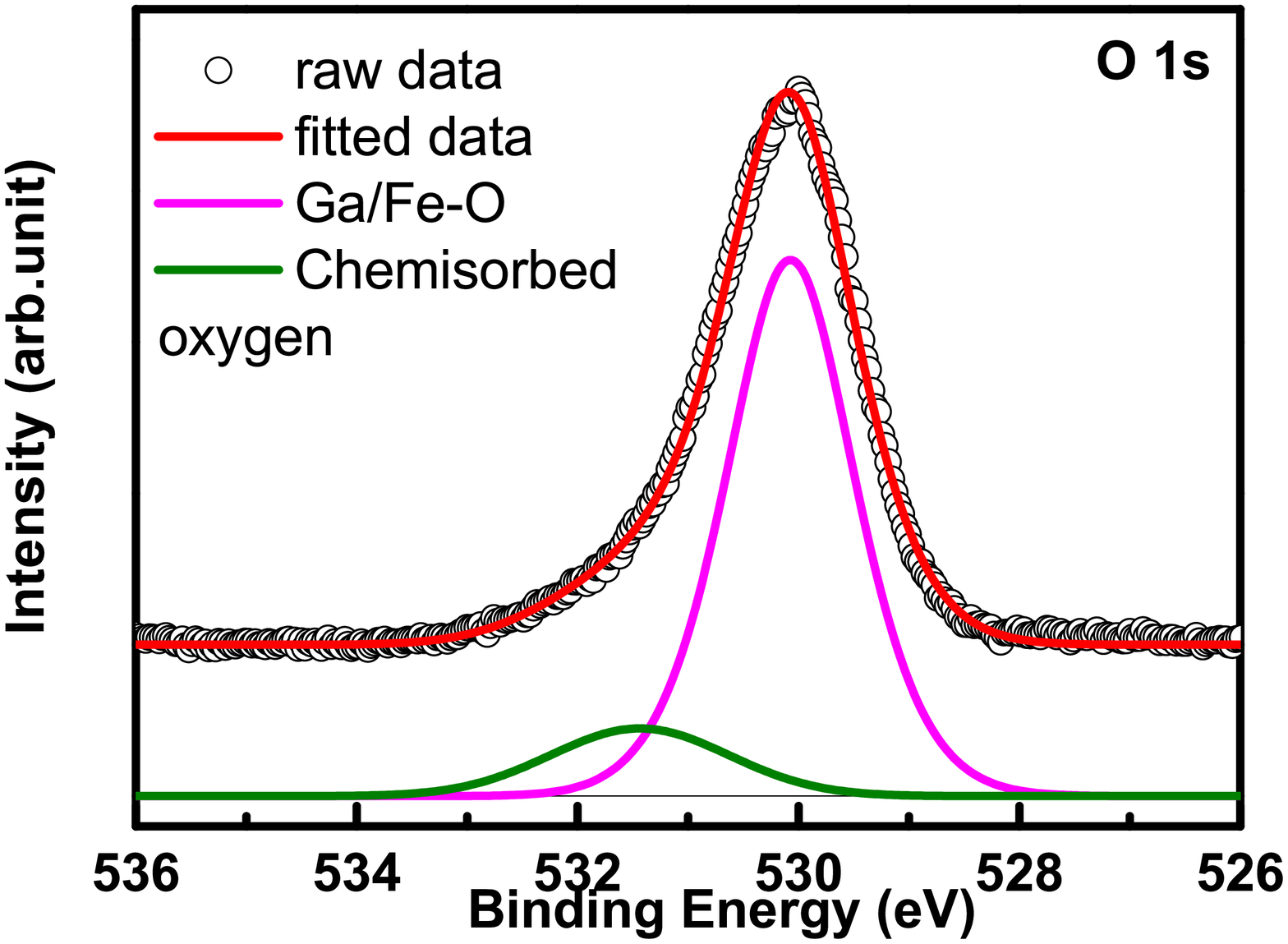}}
   \end{center}
\caption{(a) Experimental Ga 3d core-level XPS spectra (open circles) as well as the fitted spectra (red lines) for the sample are plotted together; the features around 18.7, 19.6 and 22.3 are marked as A, B and C, respectively; (b) experimental Fe 2p core-level XPS spectra and their fitting; features belonging to Fe2+, Fe3+ and satellite (sat.) are marked; (c) experimental O 1s core-level XPS spectra and their fitting. }
\end{figure}

\begin{figure}[ht!]
\centering
{\includegraphics[scale=0.25]{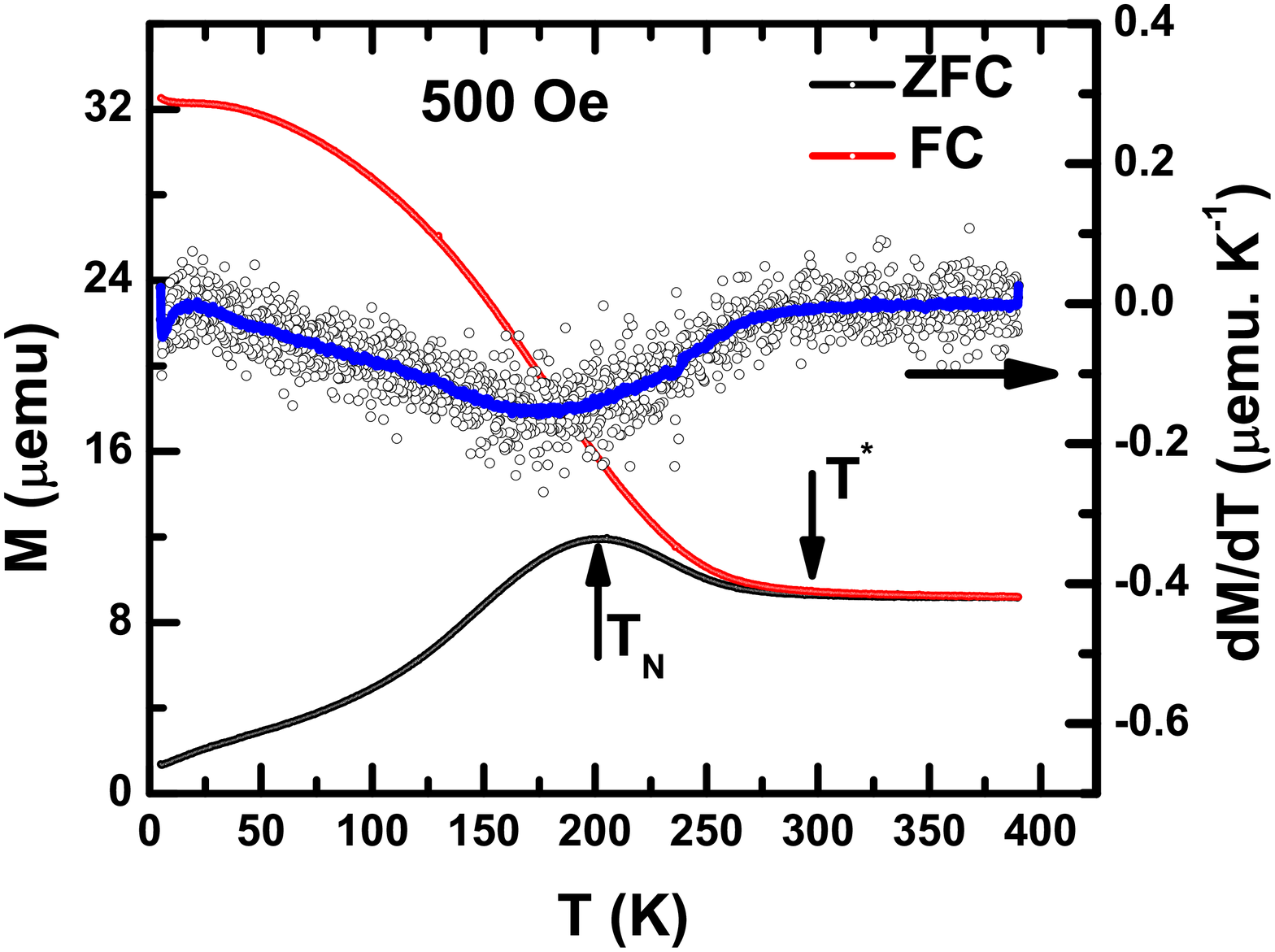}}
\caption{The zero-field-cooled (ZFC) and field-cooled (FC) magnetization ($M$) versus temperature ($T$) data with applied field $H$ $\parallel$ film surface; right axis corresponds to $dM_{FC}/dT$.}
\end{figure}

\begin{figure*}[ht!]
\begin{center}
   \subfigure[]{\includegraphics[scale=0.15]{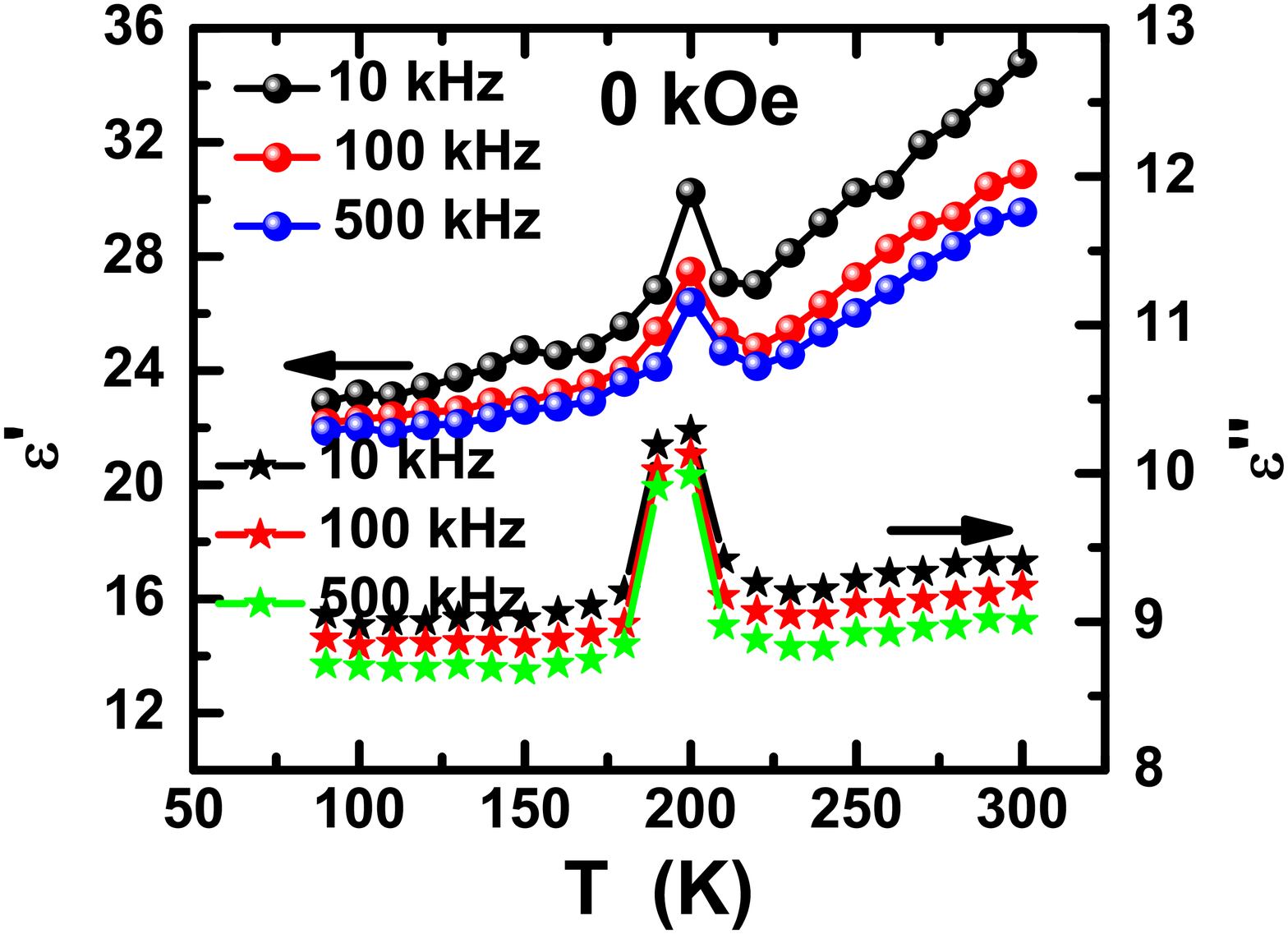}} 
   \subfigure[]{\includegraphics[scale=0.15]{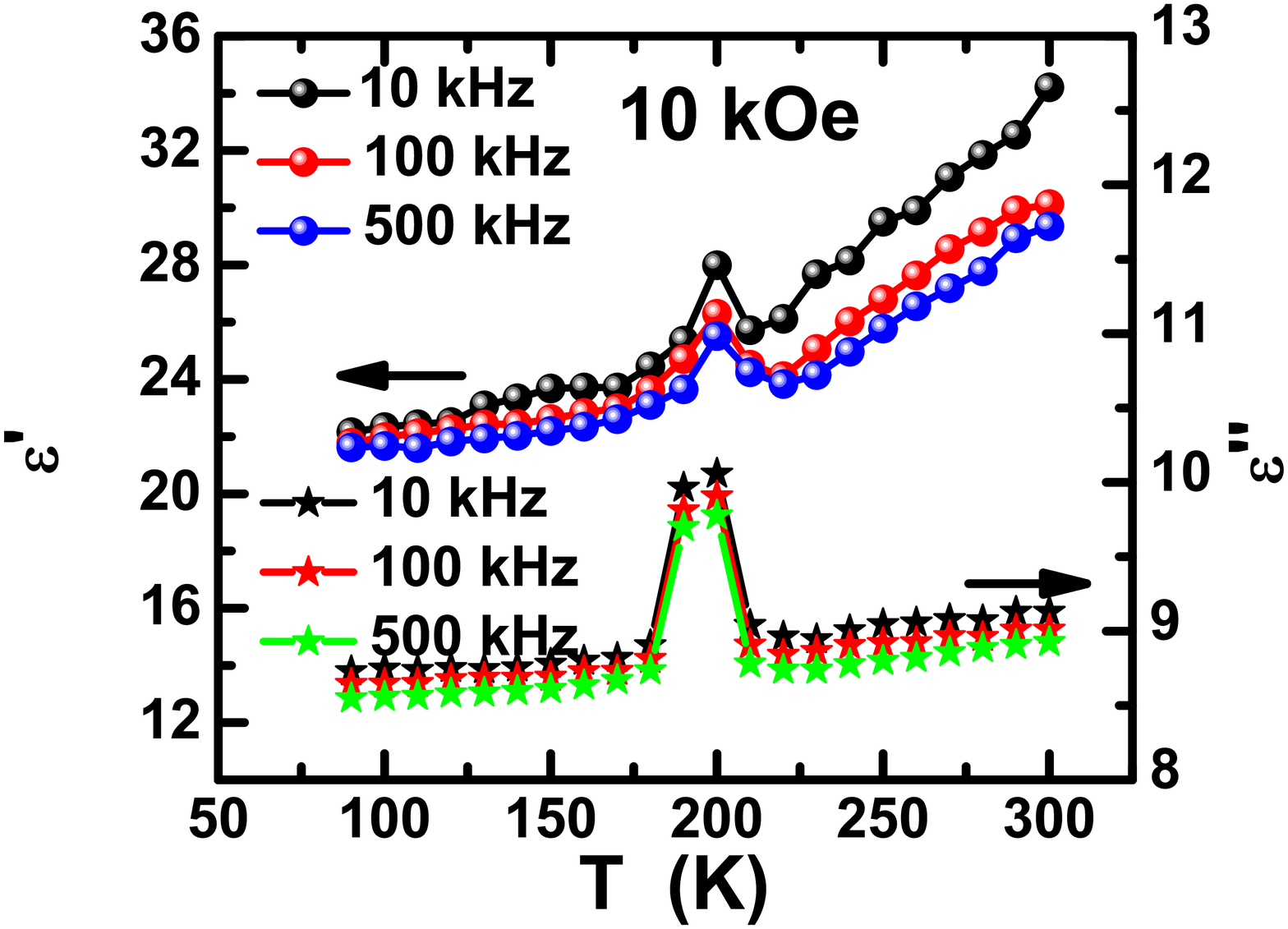}}
   \subfigure[]{\includegraphics[scale=0.15]{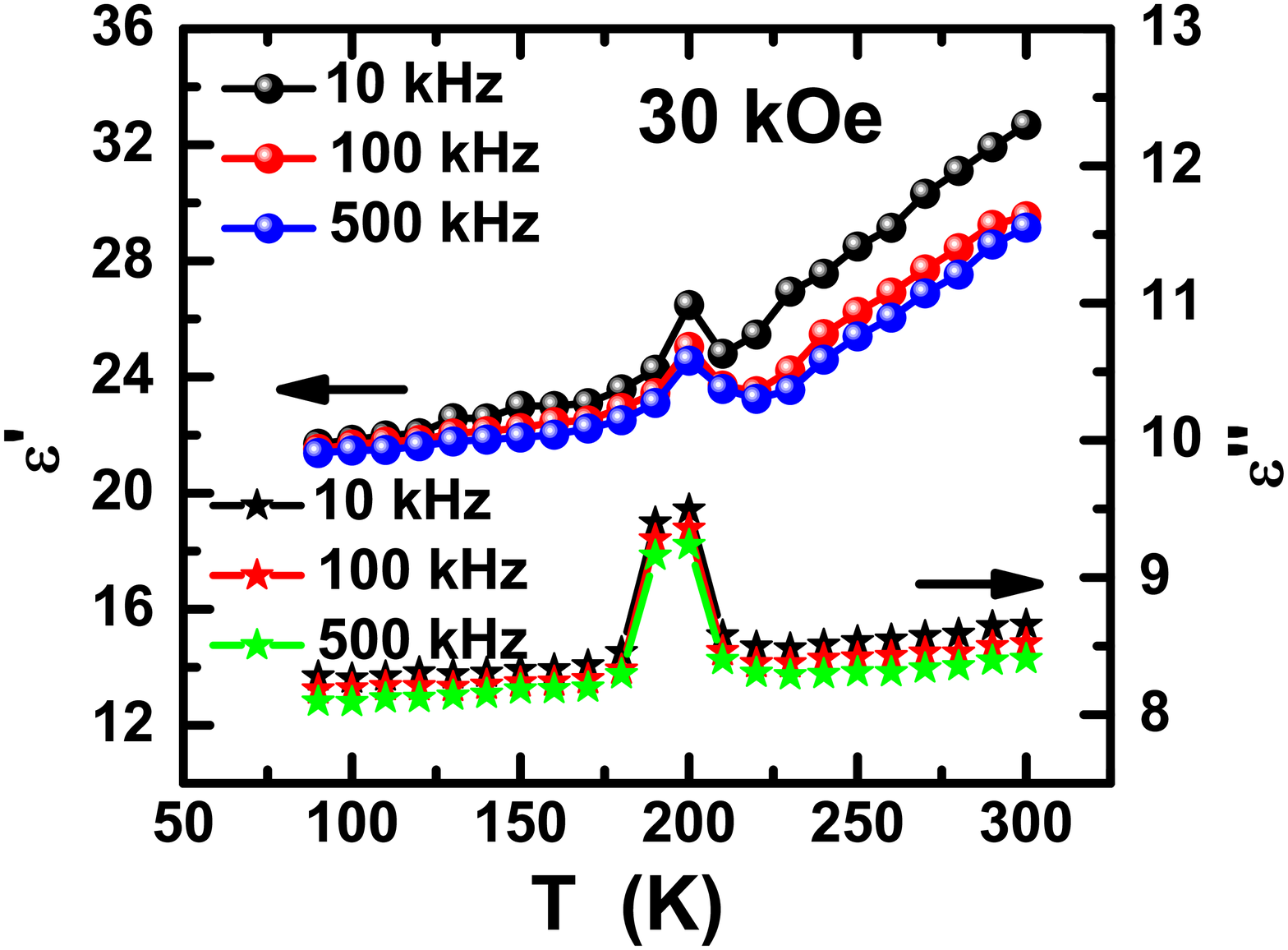}}
   \subfigure[]{\includegraphics[scale=0.15]{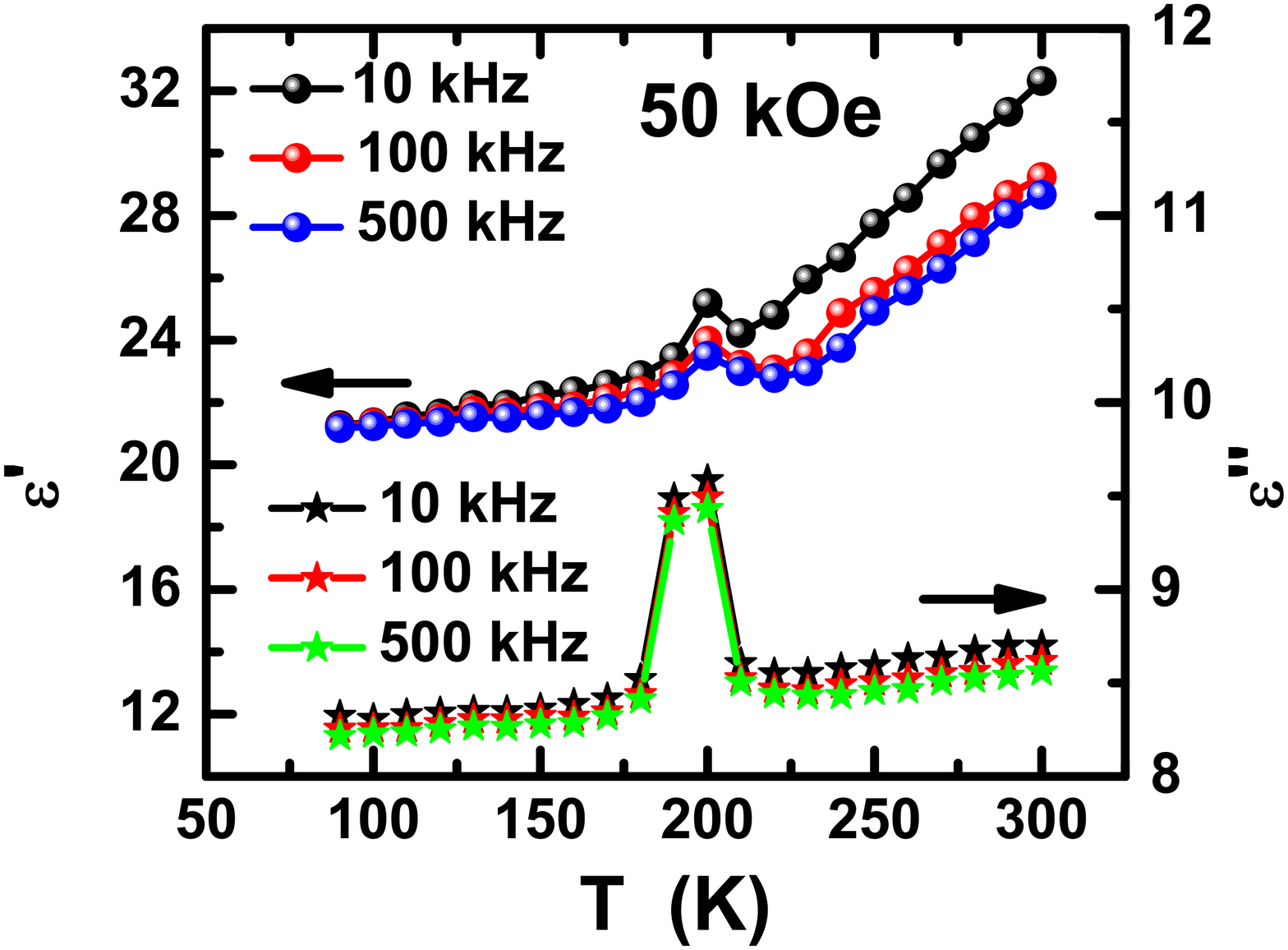}}
   \end{center}
\caption{Variation of the real and imaginary dielectric permittivity ($\epsilon '$, $\epsilon ''$) with temperature under (a) 0, (b) 10, (c) 30, and (d) 50 kOe field.}
\end{figure*}

The GaFeO$_3$ films deposited on (111)SrTiO$_3$ or (001)YSZ substrates\cite{Mukherjee,Thai,Kundaliya,Trassin,Roy,Song,Dugu,Homkar} exhibit epitaxial growth along [001] or [010] axes with nearly relaxed c- or b-axes (i.e., c or b is very close to those for the bulk sample = 9.399 \AA). The epitaxial strain - tensile or compressive depending on the substrate - along other axes varies within 0.5-2.0\%. In the present case, the out-of-plane (in-plane) strain turns out to be tensile (compressive) and comparable with the observations made by others.\cite{Mukherjee,Thai,Kundaliya,Trassin,Roy,Song,Dugu,Homkar} It may result from the anisotropic change in the bond lengths and angles due to rotation of the FeO$_6$ octahedra driven by the lattice misfit between the substrate and the film and relaxation of the strain via defects and dislocations. The anisotropic variation of the chemical strain (due to oxygen and other ion concentration variation) could also result in such anisotropy.  

Using FESEM-EDX imaging, the Ga:Fe ratio is mapped comprehensively across the macroscopic to nanoscopic scale. For example, each zone of different length scale - $\sim$3.5 mm $\times$ $\sim$2.5 mm, $\sim$100 $\mu$m $\times$ $\sim$100 $\mu$m, $\sim$20 $\mu$m $\times$ $\sim$20 $\mu$m, $\sim$3 $\mu$m $\times$ $\sim$3 $\mu$m  - has been divided into several sub-zones. The EDX spectra were collected (supplementary material) from those zones. The Ga:Fe ratio, obtained from these scanning, offers a mapping of the composition across the length scale range 10$^4$ (Fig. 2). It appears that, at the macroscopic scale, the Ga:Fe ratio varies within $\sim$1.30-1.50, i.e., the composition varies within Ga$_{2-x}$Fe$_x$O$_3$ (0.80$\le$x$\le$0.87). Inhomogeneity in the Ga:Fe ratio across the film surface, however, prevails at smaller length scales too (Fig. 2). As the frames of the Fig. 2 are scanned from right to left, it is possible to notice that with the decrease in the length scale from millimeter to micrometer to sub-micrometer, the Ga:Fe ratio first decreases and then increases again at the sub-micrometer scale. It indicates persistence of the compositional inhomogeneity (which, in turn, induces chemical strain) down to the hundreds of nanometers scale.

We used transmission electron microscopy (TEM) and high resolution TEM (HRTEM) on different (primarily, in-plane) regions of the film to map the lattice strain within a scale of a few nanometers. In Fig. 3, we show the representative TEM and HRTEM images and their analyses. The HRTEM image was analyzed by generating its fast Fourier transformation (FFT) and then its inverse (i.e. inverse FFT or IFFT) images in order to determine the lattice spacing `$d$' clearly. We could observe presence of additional planes [e.g. (221)] as different, especially in-plane, regions were exposed to the TEM. Interestingly, local variation (within a scale of few nanometers) of `$d$' could be noticed for the (221) planes. Additional data showing the local variation of `$d$' for the (130) planes are available in the supplementary material. This result provides further evidence of presence of inhomogeneity (variation in Ga:Fe ratio) driven variation in local lattice strain in the film and thus corroborates the results obtained from the mapping of Ga:Fe ratio by FESEM and EDX.

We have also examined the electronic structure of the ions using the XPS spectra of the Ga, Fe, and O. The Ga 3d spectra are very much sensitive to the introduction of Fe atoms in the lattice. The fitted spectra of the Ga 3d with three different features marked as A, B and C are shown in Fig. 4(a). The most prominent feature B is mainly due to the Ga atoms bonded with O atoms which is not affected by any kind of self-doping or coordinated with Fe atoms. We have used spin-orbit splitting 0.4 eV, branching ratio 1.5, Gaussian broadening 0.8 eV and Lorentzian broadening 0.5 eV for this fitting with an integral background. In the lower binding energy side, the feature A is attributed to the antisite defects -  mainly Ga atom on O site or O atom on Ga site and interstitial defects - Ga atoms coordinated with Ga atoms. On the other hand, the feature C at the higher binding energy side is due to the Ga atoms on Fe site (disorder). From the fitting, the relative areas of A, B and C are around 6.7, 73.3 and 19.9\%. Thus, around 20\% of the volume is due to disorder and/or oxygen vacancies. 

The Fe 2p$_{3/2}$ and Fe 2p$_{1/2}$ spin-orbit doublet peaks are at around 709.6 eV and 722.9 eV, respectively, with a pair of shake-up satellite peaks located at 6.0 eV above their spin-orbit doublet peaks [Fig. 4(b)]. The doublet peaks are wide and ascribed to the Fe-O bonds. The satellite peak is observed at 6.0 eV above Fe 2p$_{3/2}$, which confirms the dominant 2+ oxidation states of Fe. From the fitting, the relative peak areas of Fe2+ and Fe3+ states is found to be 60.0 : 37.9. 

Figure 4(c) displays the O 1s peaks fitted with two features since slight asymmetry is present in the higher binding energy side. The symmetry peak located at 530.0 eV is due to nearly stoichiometric GaFeO$_3$ lattice. Attribution of the asymmetry peak located at 531.4 eV is not straightforward. It could be partially due to chemisorbed oxygen together with oxygen vacancies present in the lattice. Therefore, the XPS data reveal both nonstoichiometry as well as disorder in the site occupancy. Presence of Fe2+ influences the remanent polarization. 
 
The zero-field cooled (ZFC) and field-cooled (FC) magnetization ($M$) versus temperature ($T$) data (Fig. 5) show onset of divergence at $\sim$300 K. The onset temperature $T^*$ of magnetic ordering, expected to vary within $\sim$200-250 K for the composition range Ga$_{2-x}$Fe$_x$O$_3$ (0.80$\le$x$\le$0.87), has actually extended to $\sim$300 K because of the presence of short-range magnetic order induced by the combined effect of epitaxial strain and compositional inhomogeneity (chemical strain). Earlier work,\cite{Arima,Remeika,Nowlin} in fact, mapped the variation of magnetic transition temperature $T_C$ with $x$. It has been shown that with the increase in $x$, the $T_C$ increases from $\sim$100 K to $\sim$350 K. In the present case, presence of finer magnetic domains has been imaged by MFM (discussed later) as well. The plot of $dM_{FC}/dT$ versus temperature (Fig. 5) indicates the transition temperature $T_N$ - which actually marks the completion of the transition - to be $\sim$200 K. The transition width, therefore, is $\sim$100 K.

\begin{figure}[ht!]
\begin{center}
   \subfigure[]{\includegraphics[scale=0.25]{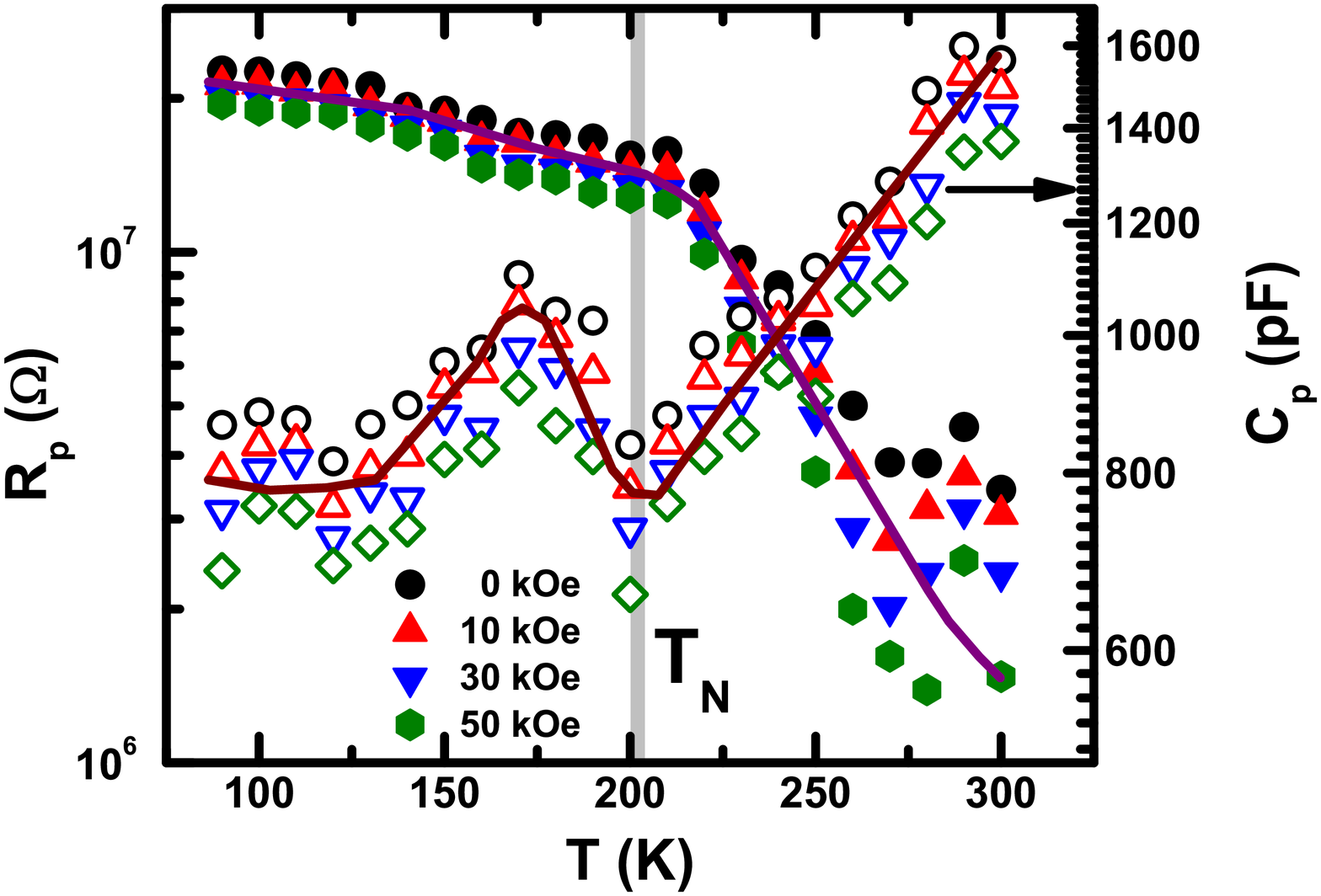}} 
   \subfigure[]{\includegraphics[scale=0.20]{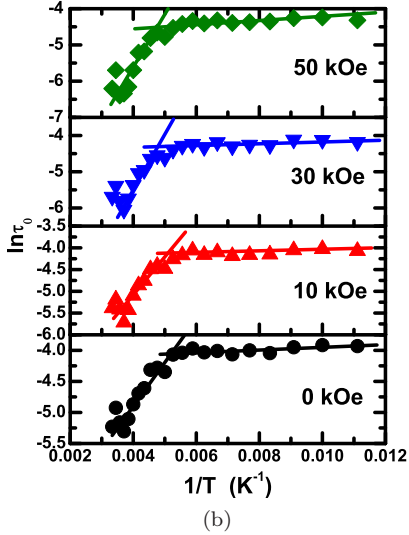}}
   \end{center}
\caption{Variation of the equivalent circuit parameters (a) $C_p$ and $R_p$, and the relaxation time scale (b) ln$\tau_0$ with temperature under different magnetic fields - 0, 10, 30, and 50 kOe.}
\end{figure}

\begin{figure}[ht!]
\centering
{\includegraphics[scale=0.25]{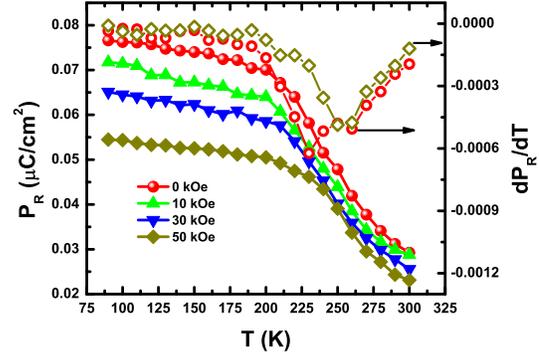}}
\caption{The variation of the remanent ferroelectric polarization with temperature under different magnetic field; right axis corresponds to the plot of $dP_R/dT$ versus temperature ($T$).}
\end{figure}

The dielectric permittivity ($\epsilon '$, $\epsilon ''$) versus temperature patterns (Fig. 6), recorded under 0-50 kOe magnetic field across 90-300 K, exhibit characteristic peak around $T_N$ due to coupling between magnetic and dielectric properties. The peak, however, is broadened over $\sim$100 K indicating the influence of the magnetic short-range order. The extent of broadening decreases under magnetic field. In spite of clear evidence of presence of magnetic short-range order, no ac-field frequency-dependent shift of $T_N$ could be observed which signifies coexistence of long- and short-range order and relatively smaller volume fraction of the short-range order.

\begin{figure}[ht!]
\begin{center}
   \subfigure[]{\includegraphics[scale=0.20]{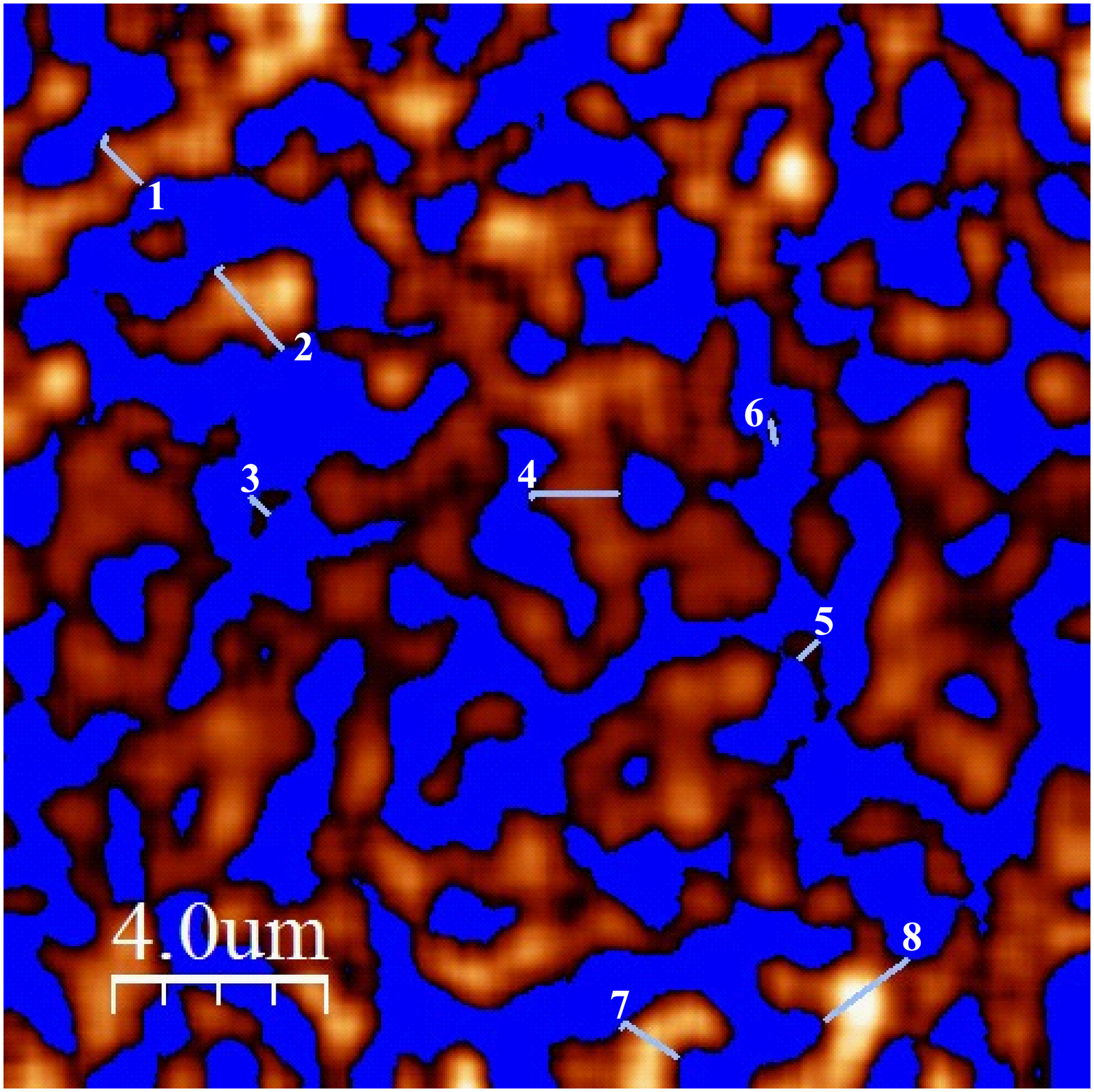}} 
   \subfigure[]{\includegraphics[scale=0.20]{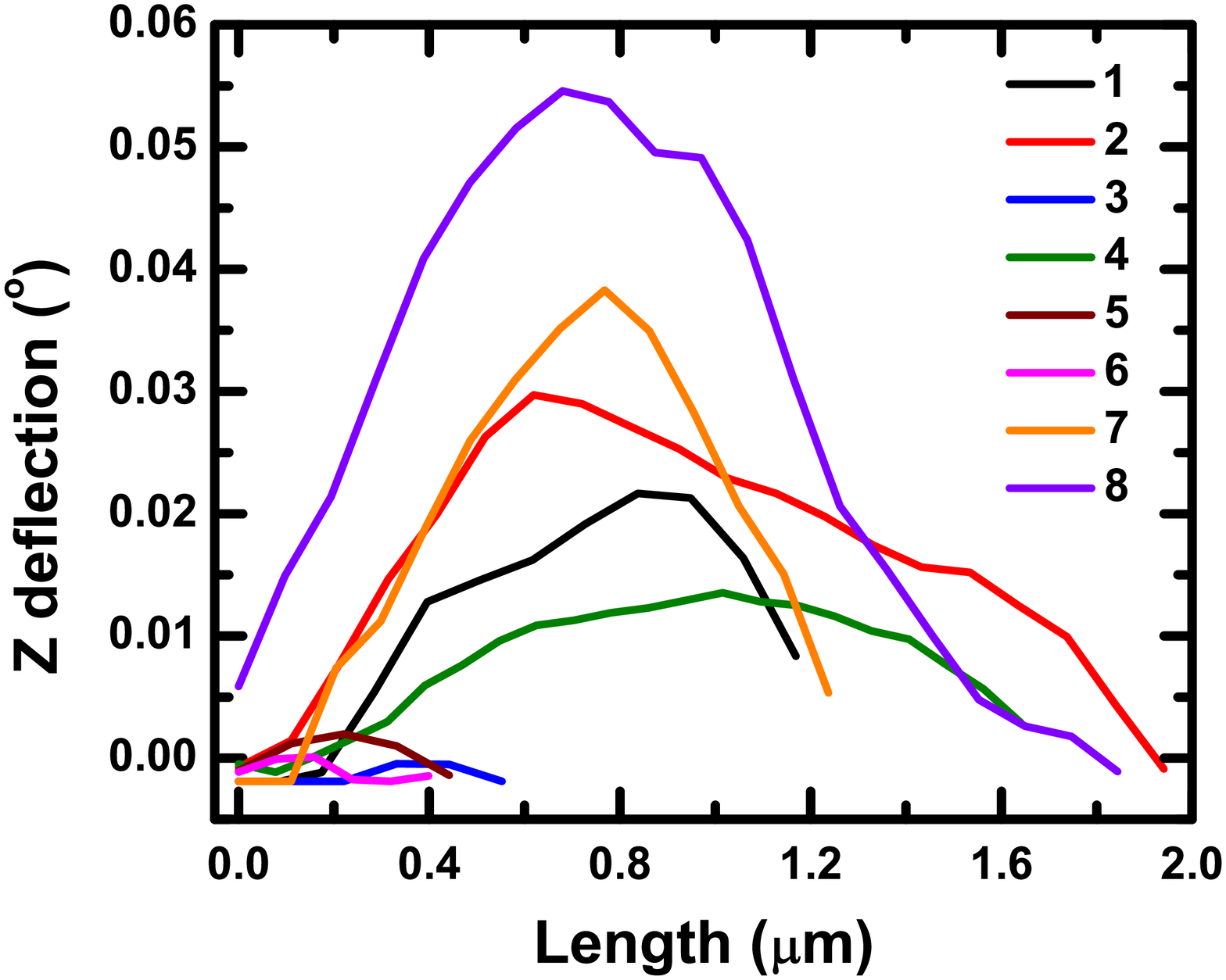}}
\end{center}
\caption{(a) The magnetic domain structure obtained from the magnetic force microscopy imaging at room temperature; (b) the line profile plots showing the size of the magnetic domains at different regions of the film; the scanned lines along which the data were recorded are shown in (a).}
\end{figure}

The dielectric relaxation spectra recorded across 100 Hz to 5 MHz at different temperature and magnetic field were fitted by Davidson-Cole model (supplementary material) which describes the non-Debye type dielectric relaxation process. The Davidson-Cole equation is given by  

\begin{center}

$Z^*$ = $R_{\infty}$ + $\frac{R_0 - R_{\infty}}{(1 + i\omega \tau_0)^{\beta}}$

\end{center}

\noindent where $R_0$ and $R_{\infty}$ are the static and high frequency resistances, respectively, $\omega$ is the frequency, $\tau_0$ is the relaxation time scale, and $\beta$ varies within 0 to 1 for non-Debye relaxation. The equivalent circuit is comprised of series connection of parallely connected resistance and capacitance sets corresponding to the bulk and interface regions of the sample. The resultant temperature and magnetic field dependence of the bulk intrinsic resistance $R_p$, capacitance $C_p$, and the relaxation time constant $\tau_0$ are shown in Fig. 7. Clear anomalous features could be observed in $R_p$, $C_p$, and $\tau_0$ around the magnetic transition temperature $T_N$. The $\tau_0-T$ patterns at below and above $T_N$ turn out to be Arrhenius $\tau_0$ = $\tau_{\infty}$exp($E/k_BT$) where $E$ is the activation energy; $E$ varies within $\sim$20-32 K below $T_N$ and within $\sim$650-1100 K at above $T_N$ for different magnetic field. Magnetic order, therefore, influences the dielectric relaxation process significantly by reducing the activation energy. However, application of magnetic field ($H$) appears to have influenced the magnetic and/or crystallographic structure subtly both at below $T_N$ and in the $T_N-T^*$ interval which leads to the enhancement of $E$ under $H$ in both the regimes.

The intrinsic remanent ferroelectric polarization ($P_R$), measured by employing a specially designed protocol\cite{Chowdhury} which helps in extracting the intrinsic switchable (i.e., hysteretic) polarization by eliminating various spurious effects, exhibits clear anomalous rise below $T_N$ due to adding up of $P_{magnetic}$ with $P_{structural}$. The $dP_R/dT-T$ plot shows (Fig. 8) shift of $T_N$ toward higher temperature and decrease in the transition width $\Delta T_N$ under $H$. Below $T_N$, of course, the $P_R$ drops (consistent with the drop in $C_p$) with the increase in $H$ ($\sim$21\% under $\sim$50 kOe at 300 K). The decrease in $P_R$ under $H$ (supplementary material) indicates field-driven subtle change in the magnetic structure and consequent decrease in $P_{magnetic}$ under $H$. Of course, the magnitude of $P_R$ is relatively smaller than the observations made in thin films or single crystals by others. This could be because of leakage due to the presence of both Fe2+ and Fe3+ ions. However, the magnetoelectric multiferroic coupling turns out to be quite strong indeed.

The magnetic domain structure has been imaged by MFM using dual pass technique in which the cantilever (commercial Co-alloy coated one of Point Probe Plus MFM Reflex coating; nominal coercivity 300 Oe) is tuned to the resonant frequency $\sim$70 kHz by digital phase-lock-loop (PLL) control system at a certain oscillation amplitude 10-50 nm. The lift height was kept constant at $\sim$120 nm. The phase shift of the cantilever due to tip-sample interaction is recorded as the MFM phase contrast image. The image is processed by the WSxM software in which the hills (red) and the holes (blue) are defined by a scale of z-deflection (scale: -0.0019$^o$ to 0.055$^o$). The processed image (Fig. 9a) shows the domain contrast clearly. The line profile analysis (Fig. 9b) shows the size of the domains to be varying within $\sim$100-2000 nm. The regions from where the line scans were captured are shown in Fig. 9a and the scale of the z-deflection (defined by the WSxM software) is used in the Fig. 9b. This image provides corroborative evidence of coexistence of finer and coarser magnetic domains and, hence, helps to understand the magnetic, dielectric, and ferroelectric properties of the film.

Above $T_N$, the strain gradient (due to compositional inhomogenity) could induce ferroelectricity due to flexoelectric effect.\cite{Deng} Near the magnetic transition, flexo effect could affect the magnetization as well.\cite{Lukashev} Influence of flexoelectromagnetic effect has recently been examined\cite{Peng} in other systems as well. Although, flexo effect could normally be observed in nanoscale thin films of different thickness where thicker films (thickness greater than the critical) could exhibit large strain gradient due to sharp relaxation of epitaxial strain via formation of defects/dislocations, in recent time, it has been shown\cite{Purohit} that flexo effect exists locally as well because of spatial variation in defect concentration. The strain gradient across a certain length scale around a defect has been mapped. Therefore, influence of flexoelectromagnetic effect is expected to be finite in the present case where local variation in Ga:Fe concentration ratio (mapped by FESEM and EDX) and consequent variation in lattice strain (mapped by TEM and HRTEM) could be observed within a length scale of few hundreds to few tens of nanometers. Since the magnetic structure in GaFeO$_3$ induces polarization\cite{Arima}, the net polarization $P$ below $T_N$ could be given by $P$ = $P_F$ + $P_{F_{flex}}$ + $P_M$ + $P_{M_{flex}}$ where $P_F$ and $P_M$ are the polarizations resulting from, respectively, structural noncentrosymmetry and magnetic structure and $P_{F_{flex}}$ and $P_{M_{flex}}$ define the polarizations due to local flexoelectromagnetic effect which gives rise to the formation of finer polar and magnetic domains. Combined influence of local flexoelectric/flexomagnetic effect induces onset of magnetic transition at room temperature (via formation of finer magnetic domains) and also influences the magnetoelectric effect. Interestingly, even though the Ga:Fe ratio is large ($>$1.0) which, according to the phase diagram reported in Ref. 3, is expected to shift the $T_N$ to lower temperature, the presence of epitaxial and chemical strain and strain gradient, in contrast, has actually shifted the onset of magnetic transition $T^*$ to higher temperature.

\section{Summary}
In summary, this work shows that even when the multiferroic orthorhombic GaFeO$_3$ film is deposited on economic and technologically important cubic Si(100) substrate by pulsed laser deposition technique, it exhibits preferential orientation (by nearly 94\%) with c-axis perpendicular to the film surface and substantial room temperature magnetoelectric coupling. Rigorous structural analysis and compositional mapping reveal coexistence of epitaxial and chemical strain and strain gradient. They, in turn, give rise to the formation of finer magnetic domains and room temperature multiferroicity. The GaFeO$_3$ film, grown on Si(100) substrate, could thus be very useful for nanospintronic applications.\\ 

\noindent $\textbf{SUPPLEMENTARY MATERIAL}$\\
\\
The supplementary material contains the XRD data for the bulk sample and epitaxial thin and also the ion positions as obtained from the Rietveld refinement of GIXRD data, the EDX spectra, additional TEM, HRTEM, and their FFT and IFFT images, remanent hysteresis loops recorded under different magnetic field, and the complex plane impedance spectra and their fitting by appropriate equivalent circuit model. It is available upon reasonable request to the author. \\
\\
\noindent $\textbf{ACKNOWLEDGMENTS}$\\
\\
The authors S.G. and S.M. contributed equally to this work. One of the authors (S.M.) acknowledges the DST-INSPIRE fellowship of Govt of India during the work.\\
\\
\noindent $\textbf{AUTHOR DECLARATION}$\\
\\
\noindent $\textbf{Conflict of Interest}$\\
\\
The authors have no conflicts to disclose.\\ 
\\
\noindent $\textbf{Data availability statement}$\\
\\
\noindent The data that support the findings of this study are available within this article and its supplementary material.

\end{document}